\documentclass[lettersize,journal]{IEEEtran}
\usepackage{amsmath,amsfonts}
\usepackage{algorithmic}
\usepackage{algorithm}
\usepackage{array}
\usepackage[caption=false,font=normalsize,labelfont=sf,textfont=sf]{subfig}
\usepackage{textcomp}
\usepackage{stfloats}
\usepackage{url}
\usepackage{verbatim}
\usepackage{graphicx}
\usepackage{cite}
\usepackage{bm}
\usepackage{enumitem}
\usepackage{booktabs}
\usepackage{multirow}
\usepackage{caption}
\usepackage[utf8]{inputenc}
\begin{document}
	
\title{Semi-supervised learning via DQN for log anomaly detection}

\author{\IEEEauthorblockN{Yingying He, Xiaobing Pei}
	\IEEEauthorblockA{\\
		\textit{Huazhong University of Science and Technology}\\
		}

	\thanks{This paper was produced by the IEEE Publication Technology Group. They are in Piscataway, NJ.}
	\thanks{Manuscript received April 19, 2021; revised August 16, 2024.}}

\markboth{Journal of \LaTeX\ Class Files,~Vol.~14, No.~8, August~2024}%
{Shell \MakeLowercase{\textit{et al.}}: A Sample Article Using IEEEtran.cls for IEEE Journals}

\IEEEpubid{0000--0000/00\$00.00~\copyright~2024 IEEE}

\maketitle

\begin{abstract}
Log anomaly detection is a critical component in modern software system security and maintenance, serving as a crucial support and basis for system monitoring, operation, and troubleshooting. It aids operations personnel in timely identification and resolution of issues. However, current methods in log anomaly detection still face challenges such as underutilization of unlabeled data, imbalance between normal and anomaly class data, and high rates of false positives and false negatives, leading to insufficient effectiveness in anomaly recognition. In this study, we propose a semi-supervised log anomaly detection method named DQNLog, which integrates deep reinforcement learning to enhance anomaly detection performance by leveraging a small amount of labeled data and large-scale unlabeled data. To address issues of imbalanced data and insufficient labeling, we design a state transition function biased towards anomalies based on cosine similarity, aiming to capture semantic-similar anomalies rather than favoring the majority class. To enhance the model's capability in learning anomalies, we devise a joint reward function that encourages the model to utilize labeled anomalies and explore unlabeled anomalies, thereby reducing false positives and false negatives. Additionally, to prevent the model from deviating from normal trajectories due to misestimation, we introduce a regularization term in the loss function to ensure the model retains prior knowledge during updates. We evaluate DQNLog on three widely used datasets, demonstrating its ability to effectively utilize large-scale unlabeled data and achieve promising results across all experimental datasets.
\end{abstract}


\section{Introduction}
With various industries continuously digitizing, the scale and complexity of software systems are gradually increasing. The likelihood of software failures is also rising, which can lead to immeasurable consequences \cite{le2022log}. Therefore, for software systems, it is particularly important during both development and operations to perform anomaly detection by analyzing system behavior patterns and resource conditions. Logs, which are semi-structured texts that record program behavior during runtime, are crucial in this regard. Compared to other core operational monitoring analysis objects such as metric data and distributed tracing data \cite{nedelkoski2019anomaly}, logs are rich in content and abundant in source, often reflecting software system anomalies \cite{zhang2022deeptralog}. By mining log data, operations engineers can proactively identify potential issues and resolve them promptly, ensuring the stable operation of systems and reducing losses caused by system failures. Currently, methods for log anomaly detection include those based on traditional machine learning, deep learning, and reinforcement learning approaches.

Traditional machine learning methods \cite{farshchi2015experience, liang2007failure, lin2016log} extract features manually from log sequences to create log count vectors based on event frequencies. These log count vectors are then used with supervised or unsupervised models to identify anomalies. While these methods are simple and require fewer computational and time resources, they struggle to handle high-dimensional log data effectively as they are based on statistical knowledge. Moreover, they overlook important information such as the temporal sequence and contextual semantics of logs\cite{han2021log}. As a result, they often have low detection accuracy and are not suitable for practical use.
\IEEEpubidadjcol

With the significant advantages of deep learning methods in feature representation learning, deep learning-based approaches, particularly recurrent neural network (RNN) models with good memory capturing capabilities\cite{du2017deeplog, meng2019loganomaly, zhang2019robust, li2020swisslog}, have taken a dominant position in the field of log anomaly detection. The unsupervised method DeepLog \cite{du2017deeplog} converts log sequences into sequential vectors and parameter value vectors, and detects anomalies by determining whether the current log deviates from the LSTM model trained from log data under normal execution. LogAnomaly \cite{meng2019loganomaly}, built upon DeepLog, introduces a similar template merging mechanism that enables the model to learn similar semantic information. By utilizing unlabeled training data, they can detect anomalies beyond specific categories. However, due to the disregard of prior label information, they often suffer from poor accuracy and generate more false positives. The supervised method LogRobust\cite{zhang2019robust} converts log sequences into semantic vectors and utilizes an attention-based Bi-LSTM model to solve the problem of log instability. However, obtaining large-scale labeled datasets is very difficult, extremely time-consuming and cost-effective. Although better performance can be achieved, it is not practical. The semi-supervised method PLELog\cite{yang2021semi} converts log sequences into sequential vectors, estimates probabilistic labels for unlabeled samples clustered based on labeled normal samples, and detects anomalies using an attention-based GRU neural network. It effectively combines the advantages of supervised and unsupervised, but it only utilizes some normal instances and completely ignores the prior information of known anomalies.

In recent years, reinforcement learning has been considered for time series anomaly detection due to its sequential decision-making capabilities\cite{pang2021toward, arshad2022deep, dong2021network}. It learns autonomously from interactions with the environment and can effectively utilize unlabeled data. QLLog \cite{duan2021qllog} constructs a Directed Acyclic Graph (DAG) from log sequences and establishes a Q-table based on the Q-learning\cite{clifton2020q} algorithm to assess whether a log execution sequence is anomalous. However, the exponential growth of the Q-table with an increasing number of log templates renders it impractical for datasets containing numerous log templates. Elaziz et al. \cite{elaziz2023deep} proposed a weakly supervised method based on deep reinforcement learning to identify anomalies in business processes using limited labeled anomaly data and abundant unlabeled data. It defines a state transition function based on Euclidean distance in the latent space of variational autoencoders to address data imbalance issues. However, it uses one-hot vector encoding to encode logs in business process trajectories, resulting in feature vectors that do not reflect semantic similarities between categories. Moreover, the data representation learned by variational autoencoders may lose important features, contributing to remaining issues of data imbalance to some extent.

In summary, the field of log anomaly detection still faces the following practical challenges:
\begin{itemize}
	\item Insufficient labeled data and underutilization of vast amounts of unlabeled data. The sheer volume of logs makes manually labeling them based on expert experience or domain knowledge nearly impossible. In real-world software systems, logs are typically not strictly labeled or unlabeled; there is often a scarcity of labeled data alongside readily available unlabeled data. While labeled data provides valuable prior information, unlabeled data may also contain anomalous patterns.
	\item  Log data categories are highly unbalanced, leading to model bias towards the majority class. In the log data acquired from software systems, the number of normal samples is often much larger than that of anomaly samples. When this highly unbalanced data is input into a model, deep learning models tend to lean more towards the majority class when learning log features. This ultimately results in the model more frequently misclassifying anomaly classes as normal classes during anomaly detection.
	\item The ability to identify anomalies is poor, leading to many false negatives and false positives. Anomalies are often unpredictable, they could be caused by a failure in some underlying component of the system or by a certain type of network attack. In practical applications, it is difficult to obtain a training set that covers all possible anomalies. As a result, the trained model may have a poor ability to recognize anomalies, frequently resulting in false negatives and false positives.
\end{itemize}

To solve the above problems, in this paper, we propose a deep reinforcement learning-based method (DQNLog) for software log anomaly detection. The aim is to utilize a small number of labeled samples while exploring a large number of unlabeled samples to improve the performance of anomaly detection. The method first utilizes pre-trained language models to transform log data into numerical vectors with semantic features, based on log parsing. Subsequently, with log anomaly detection as the specific objective, key components of the deep reinforcement learning model are optimized to enhance the learning and training process of the agent. In particular, the data imbalance issue is mitigated through a state transition function that favors anomalies, and the agent's behavior is evaluated using a joint reward function sensitive to anomalies. This incentivizes the agent to effectively utilize labeled prior information and potential information from unlabeled sets, thereby reducing both false negatives and false positives to improve the model's detection performance. Our work makes the following contributions:

\begin{itemize}
	\item To address the issues of insufficient labeled data and underutilization of large-scale unlabeled data in current software systems, we optimized the key components of the deep reinforcement learning model and developed an agent based on deep Q-network to achieve log anomaly detection. During the training process, the agent does not confine the search for anomalies to the given labeled examples but autonomously interacts with an environment constructed from a small set of labeled samples and a large-scale unlabeled sample set to learn labeled anomalies and actively explore potential anomalies in the unlabeled set.
	\item To address the issue of highly imbalanced log data, we designed a state transition function biased towards anomalies based on cosine similarity. The environment returns the next state that is most likely to be an anomaly according to the agent's behavior. This ensures that the model does not frequently favor the majority class in imbalanced datasets but rather tends to identify anomaly log sequences that contain more informative patterns.
	\item To address the issue of knowledge forgetting during the training process of the agent, we introduce a regularization term into the loss function, constructing a regularized loss function. This allows the agent to directly utilize known prior information during parameter updates, preventing the agent from deviating from the normal trajectory due to potentially inaccurate estimates based solely on the neural network output. This ensures that the updates of the agent are minimally affected by the unconverged network values.
	\item Experimental results demonstrate that, compared to the baseline methods, it effectively utilizes a small amount of labeled data and large-scale unlabeled data, achieving superior detection performance.
\end{itemize}
\section{Log Terminology}
A \textbf{log message} refers to the textual statement output during the software runtime, which includes log event and log parameters. Log events are statements with specific meanings that typically summarize the events expressed in the logs\cite{zhou2022deepsyslog}. Log parameters typically refer to constant values, IP addresses, task IDs (such as $block\_id$
 in HDFS logs and $node\_id$ in BGL logs), file names, etc. Log parsing is the process of extracting log events and log parameters from log messages. A log sequence consists of a series of log events partitioned by strategies such as task IDs or sliding windows. If one log message in a log sequence is anomalous, the entire log sequence is considered anomalous; otherwise, it is normal. Each log event is transformed into a semantic vector through semantic embedding, denoted as $V\in\mathbb{R}^d$, where $d$ represents the dimensionality of the semantic vector. The sequence composed of vectorized log events is referred to as a semantic vector sequence, denoted as $s=\left[V_1,V_2,\cdots,V_T\right]$, where $T$ is the number of log events in the log sequence.

\section{Proposed approach}
\subsection{Overview}
In this paper, we propose a deep reinforcement learning approach for software log anomaly detection. In this method, raw log data is transformed into fixed-dimensional semantic vectors, and leveraging the autonomous learning and decision-making capabilities of deep reinforcement learning, the model effectively utilizes a small amount of labeled data and large-scale unlabeled data to enhance the performance of log anomaly detection.

\begin{figure}[!t]
	\centering
	\includegraphics[width=3.5in]{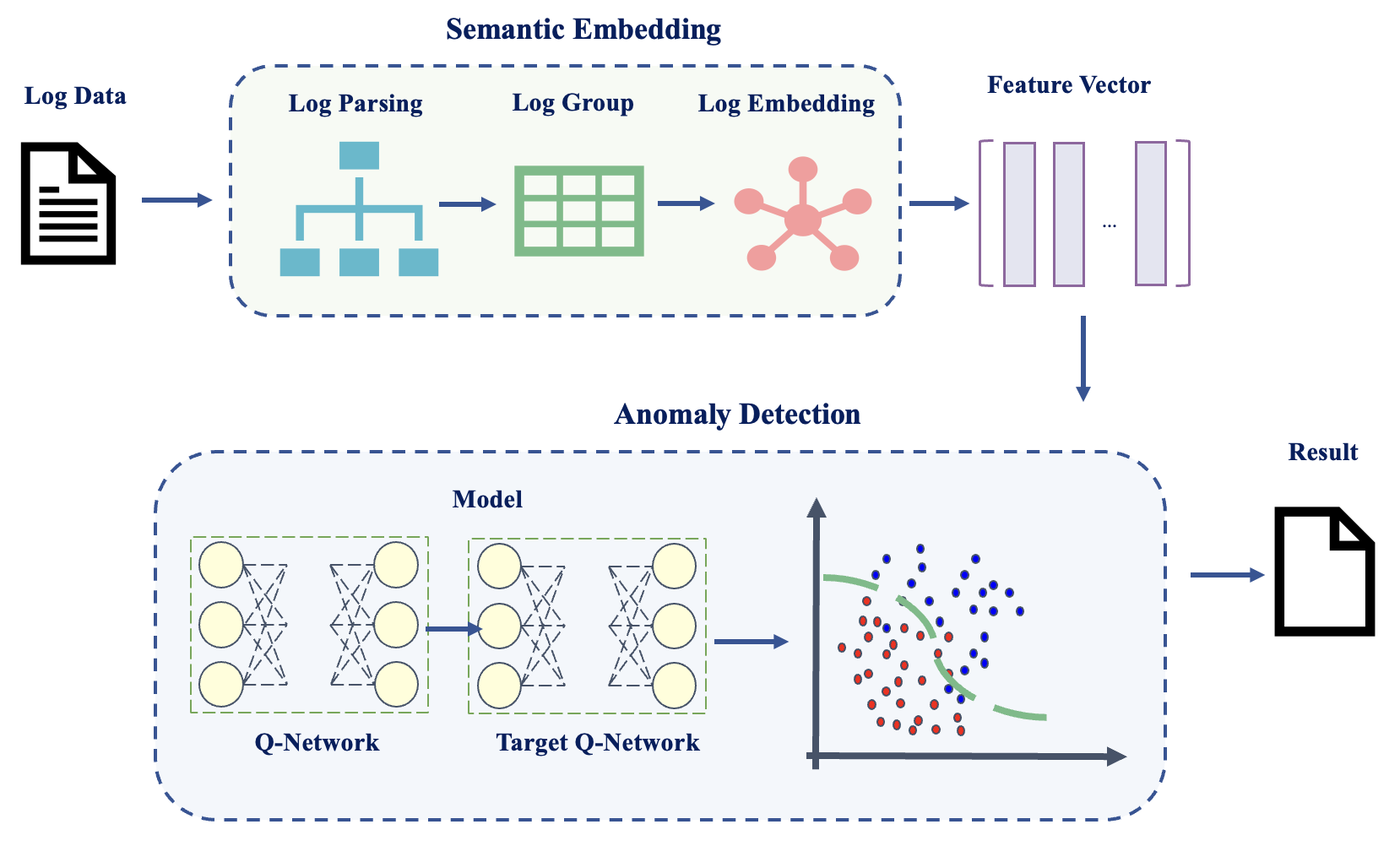}
	\caption{Overview of DQNLog}
	\captionsetup{justification=centering}
	\label{overview}
\end{figure}

Figure \ref{overview} presents an overview of DQNLog, which comprises two stages: semantic embedding and anomaly detection model building. In the semantic embedding stage, the raw log data undergoes log parsing to extract log events. These log events are then grouped into log sequences. Afterward, the log sequences are subjected to semantic embedding to obtain log semantic vectors. Moving on to the anomaly detection stage, a model is first constructed. Subsequently, the model is trained using the log semantic vectors, and the effectiveness of anomaly detection is evaluated.

\subsection{Semantic Embedding}
\begin{figure}[!t]
	\centering
	\includegraphics[width=4in]{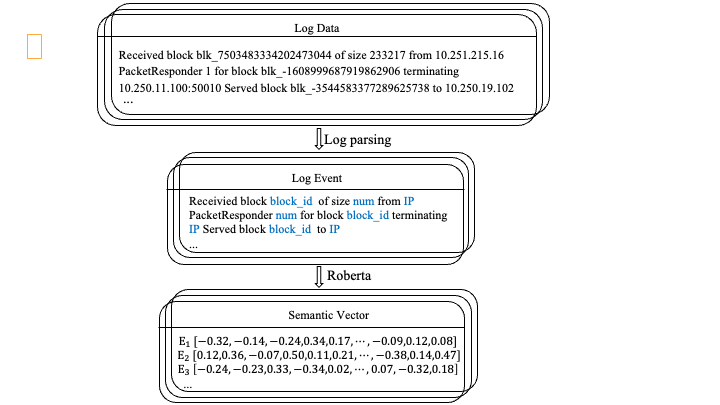}
	\caption{The process of log semantic embedding}
	\captionsetup{justification=centering}
	\label{semantic}
\end{figure}

The log data generated by software systems is unstructured text, akin to natural language data, which machines cannot inherently recognize or understand. Therefore, before performing anomaly detection, we first transform the log data into fixed-format feature vectors. The process of log semantic embedding involves log parsing, log grouping, and log embedding, as depicted in Figure \ref{semantic}.
\subsubsection{Log Parsing}
Log data is semi-structured text composed of static constants that remain unchanged and dynamic variables that vary with runtime behavior\cite{li2023did}. Extracting common segments and unique segments from raw log data is typically a preprocessing step for downstream log tasks\cite{tao2023biglog}. Due to its adaptive parsing capability for various types and formats of log messages, Drain\cite{he2017drain} method demonstrates excellent performance even across different datasets. Therefore, we employ the Drain method to extract templates from raw log data.

\subsubsection{Log Grouping}
Since logs are sequentially output by software systems based on the occurrence of events, adjacent logs exhibit correlations. Therefore, for subsequent log analysis tasks, it is common practice to input logs within a certain time frame \cite{sedki2022effective}. In our work, for HDFS and BGL datasets that include session identifiers, we group logs based on session IDs using the session window strategy, which better reflects the logical flow of software execution and the sequence of events. Conversely, for the Thunderbird dataset, we employ the sliding window strategy to capture changing trends and periodic patterns in system behavior.

\subsubsection{Log Embedding}
Log embedding maps different log templates into a vector space, providing effective feature representation and data processing for downstream anomaly detection tasks. High-quality log feature extraction should capture semantic information from log messages, including word semantics and contextual information. DQNLog utilizes the pretrained language model Roberta to obtain semantic vectors of logs. Roberta \cite{liu2019roberta} is an enhanced version of BERT\cite{devlin2018bert}, employing dynamic masking strategies, larger training datasets, and increased batch sizes to enhance data diversity, improve model understanding, and boost generalization capabilities, resulting in more accurate semantic vectors.
\subsection{DRL Tailored for Log Anomaly Detection}
\subsubsection{Modeling of Markov Decision Processes}
To fully utilize the small labeled dataset and the large unlabeled dataset, the log anomaly detection problem is first modeled as a Markov Decision Process\cite{arulkumaran2017deep}. By defining key components such as state, agent, action, environment, and reward\cite{luo2024survey}, deep reinforcement learning can be effectively applied to the field of log anomaly detection.
\begin{itemize}[leftmargin=*]
	\item State: The state space $\mathcal{S}=\mathcal{D}$, where each state $s\in\mathcal{S}$ represents a log semantic sequence.The size of $\mathcal{S}$ is the same as the size of the training set $\mathcal{D}$, which includes a limited labeled dataset and a large unlabeled dataset.
	\item Action: The action space 
		$\mathcal{A}=\left\{a^0,a^1\right\}$, where $
		a^0$ and $a^1$ correspond to the actions `normal' and `anomaly'.
	\item Agent: Implemented using an attention-based Bidirectional Long Short-Term Memory (Bi-LSTM) neural network, it seeks an optimal action from the action space $	\mathcal{A}$ for the current state.
	\item Environment:  Simulates the agent's interaction, accomplished through a state transition function $g\left(s_{t+1}|s_t,a_t\right)$ that is biased towards anomalies.
	\item 	Rewards: The reward function is defined as $r\left(s_t,a_t\right)=h\left(r^e|s_t,a_t\right)+f\left(r^i|s_t\right)$, where $h\left(r^e|s_t,a_t\right)$ represents the extrinsic reward given by the environment based on predicted and actual labels, and $f\left(r^i|s_t\right)$ denotes the intrinsic reward provided by the exploration mechanism to measure novelty.
\end{itemize}
\subsubsection{Model framework based on DRL}
The model based on deep reinforcement learning aims to fully utilize the small labeled dataset $\mathcal{D}_l$ and effectively explore the large unlabeled dataset $\mathcal{D}_u$ to improve the performance of log anomaly detection. It achieves goal by approximating the optimal action-value function (Q-value function), which refers to the expected return when the agent takes action a in state s under policy $\pi$. It evaluates the goodness or badness of taking action $a_t\in\mathcal{A}$ in the current given state $s_t\in\mathcal{S}$ under policy $\pi$. Its definition is as Equation (1).
 \begin{equation}
	Q^\pi\left(s,a\right)=\mathbb{E}\left[\sum_{t=0}^{T-1}{\gamma^tr_{t+1}}|s_t=s,a_t=a\right]\label{equ1}
\end{equation}

Where $\gamma$ is the discount factor, representing the value of future rewards relative to current rewards. Typically, $\gamma \in [0,1]$, indicating that future rewards are considered less valuable than current rewards. Under policy $\pi$, the total expected return at the current time step $t$ is the sum of the immediate reward and the discounted future rewards.

The optimal policy 
\begin{math}
	\pi^*
\end{math}
has an expected return that is greater than or equal to all other policies. It can be achieved by greedily selecting actions that maximize the given action-value function 
\begin{math}
	Q^\pi\left(s,a\right)
\end{math}
for each state, i.e., 
\begin{math}
	\pi(s)=\arg\max_a{Q^\pi(s,a)}
\end{math}. When this update equation converges, the action-value function becomes consistent across all states, resulting in the optimal policy 
\begin{math}
	\pi^* = \arg\max Q^*(s,a)
\end{math}
and the corresponding optimal action-value function is shown in Equation (2).
\begin{equation}
	Q^\ast\left(s,a\right)=\max_\pi{\mathbb{E}\left[U_t|s_t=s,a_t=a,\pi\right]}
	\label{equ2}
\end{equation}

\begin{figure}[!t]
	\centering
	\includegraphics[width=3.5in]{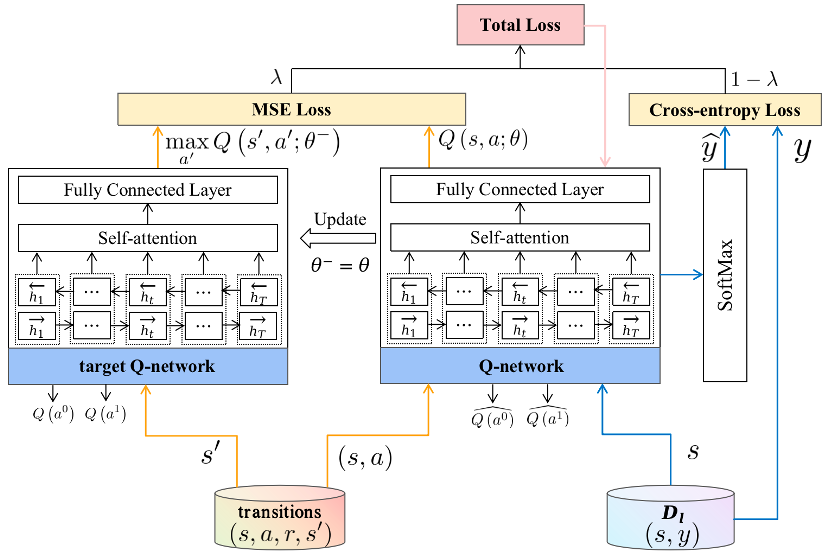}
	\caption{The overall framework of the model}
	\label{framework}
\end{figure}

To obtain the optimal action-value function under the optimal policy $\pi^*$, we employ the value-based DQN\cite{mnih2015human} algorithm, which introduces Q-network and a delayed update target Q-network. By using the neural network $Q(s,a;\theta)$ to approximate $Q^* (s,a)$ and applying experience replay\cite{zhang2017energy}, we define the overall model framework as shown in Figure \ref{framework}.

The proposed framework’s goal is to train the Q-network by interacting with the environment created from training data. During training, the agent receives experience samples from the stored experience and labeled samples from the labeled dataset. On one hand, it updates the parameters of the agent based on the difference between the current Q-value and the target Q-value predicted by the two networks. On the other hand, an additional cross-entropy loss term is introduced to constrain the agent's optimization not only based on the estimation of the Q-value function but also on the prior label information.

Both the Q-network and the target Q-network take log semantic vector sequences as input and output the corresponding Q-values for the log sequences classified as anomaly or normal. When improving its strategy, the agent receives transitions $(s_t,a_t,r_t,s_{t+1})$ from the stored experience. It uses the Q-network's output for $a_t$ in $s_t$ as the current Q-value and the maximum Q-value from the target Q-network for $s_{t+1}$ as the target Q-value, aiming to minimize the mean squared error loss between them. Simultaneously, the agent uses supervision signals from prior labels in $\mathcal{D}_l$. It applies softmax to the Q-values for all actions in labeled sequences from $\mathcal{D}_l$ to get predicted classifications, minimizing a regularization term based on these predictions and the prior labels.

In each iteration, the update of the agent's parameters is guided by the two loss functions, and the model is considered converged when the parameters of the Q-network and the target Q-network are consistent. The final obtained model is used for log anomaly detection, where the input is a log semantic vector sequence to be detected, and the output with the maximum Q-value is considered as the final classification result.

\subsubsection{ Components of the model based on DRL}
\begin{figure}[!t]
	\centering
	\includegraphics[width=3.5in]{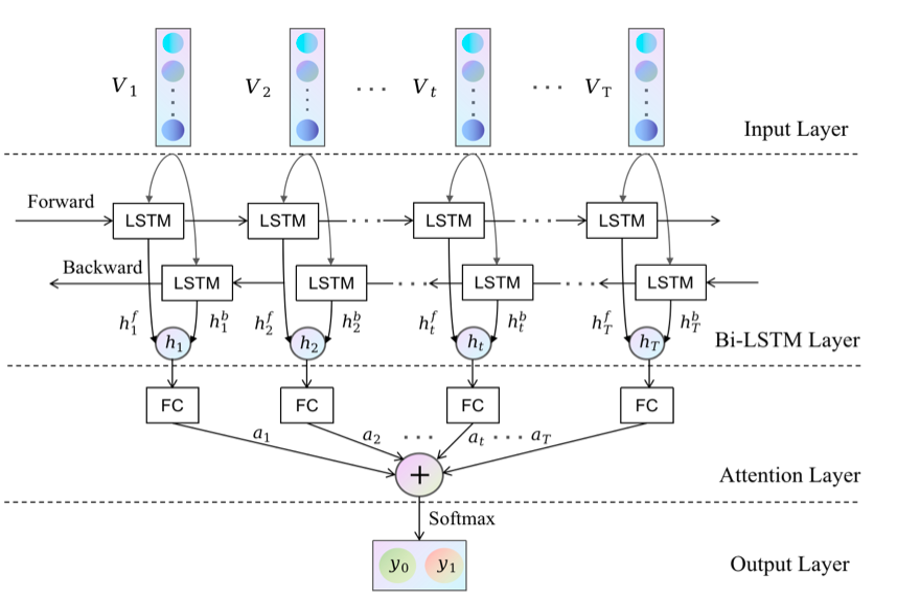}
	\caption{The specific architecture of the agent}
	\label{bilstm}
\end{figure}

In the deep reinforcement learning model, the agent and the environment are core components. The agent learns how to take actions to achieve specific goals through interaction with the environment. The design of each key component in the model is as follows:
\paragraph{Agent}
The agent is the main entity performing learning tasks. It perceives states and determines the optimal policy by learning the optimal action-value function that can evaluate the value of the action executed in the given state. Considering the characteristics of log data, such as temporality and instability, the attention-based Bi-LSTM neural network is used as the core of the agent to approximate the Q-value function. The specific architecture of the agent is illustrated in Figure \ref{bilstm}, comprising four network layers: an input layer, Bi-LSTM network layer, self-attention layer, and fully connected layer. It accepts semantic vectors of log data as input and outputs two estimated Q-values corresponding to taking normal and anomalous actions.

Bi-LSTM, an extension of LSTM, effectively captures bidirectional information by splitting the hidden neuron layer into forward and backward passes. The hidden states 
\begin{math}
	h_t^f
\end{math} and 
\begin{math}
	h_t^b
\end{math} from both directions are concatenated to form the combined hidden state 
\begin{math}
	h_t
\end{math}. To mitigate the influence of noise, an attention mechanism is introduced, assigning varying weights to log events. Implemented as a fully connected layer, the attention layer computes the attention weight 
\begin{math}
	a_t
\end{math} for each log event. The hidden states are subsequently weighted and summed based on their respective attention weights. Finally, the resulting vector undergoes classification through a softmax layer.

The attention-based Bi-LSTM model can capture long-term dependencies within log sequences and autonomously discern the importance of individual log events, thereby achieving good performance on unstable logs. Throughout the learning process, the agent continually minimizes the defined regularized loss function to update the parameters of the Q-network, until the updated parameters align with those of the target Q-network, ultimately leading to a fully trained agent.

\paragraph{Transition function biased towards anomalies}
The agent improves its anomaly detection capability by learning from the states returned by the environment. The transformation of environmental states is determined by the state transition function $g(s_{t+1} |s_t,a_t )$, which significantly influences the agent's ability to identify anomalies. In log data, the majority of data belongs to the normal class, with only a small portion classified as anomalies, resulting in a severe class imbalance. This imbalance can cause the model to heavily favor the majority "normal class," leading to weaker recognition of anomalies.

To alleviate this class imbalance problem, we propose a state transition function biased towards anomalies. This allows the agent to receive more states that are potentially anomalous, thereby learning more features of anomalies and enhancing its ability to recognize them.

Considering that the defined state space includes the labeled set $\mathcal{D}_l$ and the unlabeled set $\mathcal{D}_u$, the designed state transition function consists of two parts: the function $g_l$ for exploiting $\mathcal{D}_l$, and the function $g_u$ for exploring $\mathcal{D}_u$. 

In $g_l$, since the sample size of $\mathcal{D}_l$ is relatively small overall,the environment randomly selects the next state $s_{t+1}$ from dataset $\mathcal{D}_l$ for the current state $s_t$, regardless of the agent's action, enabling the agent to equally exploit the labeled samples in $\mathcal{D}_l$.

In $g_u$, due to the larger scale of the $\mathcal{D}_u$ sample size, the environment biases the sampling of the next state $s_{t+1}$ towards anomalies from $\mathcal{D}_u$ by considering both the agent's action $a_t$ and the cosine similarity to $s_t$.This acceleration expedites the agent's search for anomalies in large-scale imbalanced datasets, enhancing its efficiency in exploring $\mathcal{D}_u$. The definition of $g_u$ is shown in Equation (3).
\begin{equation}
	g_u\left(s_{t+1}|s_t,a_t\right)=\begin{cases}
		\arg\min_{s\in S}{d_{cos}\left(s_t,s\right)}, \quad \text{if }  a_t=a^1 \\
		\arg\max_{s\in S}{d_{cos}\left(s_t,s\right)},\quad \text{if } a_t=a^0
	\end{cases}
	\label{equ3}
\end{equation}
Where $S$ is a random subset of $\mathcal{D}_u$, $d_{cos}\in[-1,1]$ denotes cosine similarity between $s_t$ and $s$, cosine similarity is ideal for semantic vector comparison due to its normalization and directional consideration. $d_{\text{cos}}$ closer to 1 indicates greater similarity between $s_t$ and $s$, while closer to -1 signifies more dissimilarity. If the current state $s_t$ is deemed anomalous (action $ a^1 $), $g_u$ selects $s\in S$ closest to $s_t$ as the next state $s_{t+1}$; otherwise, it picks the most dissimilar $s$ as $ s_{t+1}$.

To ensure a balance between exploitation and exploration, the probabilities of using both $g_l$ and $g_u$ are set to 0.5, allowing both to be equally utilized by the agent.
\paragraph{Joint reward function}
The reward function assesses the quality of the agent's actions in the environment and guides policy updates. The reward value is feedback provided by the reward function based on the actions taken by the agent. Reasonable reward values can drive the agent to learn effective strategies and achieve task goals more quickly. However, unreasonable reward values can lead the agent into learning pitfalls, resulting in behaviors contradictory to the task goals. In the field of anomaly detection, false positives and false negatives are common. Designing a reasonable reward mechanism can effectively mitigate false positives and false negatives, enabling the agent to learn more about anomalies.

We propose a joint reward function based on an external reward function $h(s_t,a_t)$ and an intrinsic reward function $f(s_t)$ to drive the agent to effectively utilize labeled datasets and actively explore unlabeled datasets, thereby avoiding false positives and false negatives. 

\begin{itemize}
	\item External reward function $h$
\end{itemize}

The external reward measures the performance of the agent in detecting labeled anomalies and provides a reward signal $r^e$ for the action $a_t$ taken by the agent in state $s_t$. The external reward function h provides different rewards based on the different datasets from which the input state $s_t$ originates, as defined in Equation (4).
\begin{equation}
	r_t^e=\ h\left(s_t,a_t\right)=\begin{cases}
		r_1&\quad \text{ for TP},s_t\in\mathcal{D}_l \\
		r_2&\quad \text{ for TN},s_t\in\mathcal{D}_l \\
		-r_3&\quad \text{ for FP},s_t\in\mathcal{D}_l \\
		-r_4&\quad \text{ for FN},s_t\in\mathcal{D}_l \\
		0&\quad \text{if } a_t=a^0,s_t\in\mathcal{D}_u \\
		-1&\quad \text{if } a_t=a^1,s_t\in\mathcal{D}_u 
	\end{cases}
	\label{equ4}
\end{equation}

For $s_t\in \mathcal{D}_l$, the determination of $r^e$ is directly based on the comparison between the predicted labels and the actual labels, with $r_1,r_2,r_3,r_4>0$. TP denotes correct anomaly recognition by the agent, TN represents correct normal recognition, FP signifies false anomaly recognition, and FN indicates false normal recognition of an anomaly by the agent.

In the field of anomaly detection in logs, since the proportion of normal samples far exceeds that of anomalous samples, it is common for the agent to correctly identify normal samples. In comparison, correctly identifying anomalies is more valuable. Therefore, a higher reward value is set for TP compared to TN, i.e., $r_1>r_2$. Additionally, ignoring actual anomalies often poses a greater risk than generating false alarms. Consequently, compared to FP, a stricter penalty is imposed for FN, i.e., $r_4>r_3$. In this paper, we set $r_1=1, r_2=0.1, r_3=0.4, r_4=1.5$.

For $s_t\in \mathcal{D}_u$, the intelligent agent maintains a neutral attitude. Although $\mathcal{D}_u$ is unlabeled, the majority of its data is normal. Therefore, for the vast majority of normal predictions made by the intelligent agent, no reward is given, i.e., the reward value is 0. At this point, the final reward depends solely on the intrinsic reward function. Conversely, the intelligent agent incurs negative rewards as penalties to avoid excessive reliance on inherently uncertain anomalous samples.
\begin{itemize}
	\item Intrinsic reward function $f$
\end{itemize}

The intrinsic reward is a reward provided by the exploration mechanism, which gives a reward signal $r^i$ to the agent for taking action $a_t$ in state $s_t$, encouraging the agent to explore potential anomalies within the unlabeled dataset $\mathcal{D}_u$. The determination of the value of $r^i$ is based on the evaluation of a pre-trained LogRobust\cite{zhang2019robust} model, which is defined as shown in Equation (5).
 \begin{equation}
	r_t^i=f\left(s_t\right)=\begin{cases}
		1-\frac{rob_p}{\delta},\quad rob_p<\delta \\
		\left(\frac{rob_p-\delta}{1-\delta}\right)-1,\quad rob_p\geq \delta
	\end{cases}
	\label{equ5}
\end{equation}
Where ${rob}_p\in\left[0,1\right]$ denotes the probability predicted by the LogRobust model for the normalcy of the current state $s_t$, and $\delta$ is the specified confidence threshold.If ${rob}_p<\delta$, $s_t$ is deemed abnormal, and $r_t^i$ falls within $(0, 1]$. Conversely, if ${rob}_p\geq\delta$, $s_t$ is considered normal, and $r_t^i$ falls within $[-1, 0]$.When ${rob}_p$ approaches 0, $s_t$ is highly abnormal, prompting a high reward to encourage exploration of novel states. Conversely, if ${rob}_p$ approaches 1, $s_t$ is confidently normal, and no reward is given to prevent biased exploration of normal states by the agent.

 In order to achieve the agent's exploitation of $\mathcal{D}_l$ and the exploration of $\mathcal{D}_u$, we define the final reward as Equation (6).
\begin{equation}
	r=r^i+r^e
	\label{equ6}
\end{equation}
\subsection{Regularized loss function}
When making decisions, the original DQN agent approximates the Q-value function using a deep neural network and always selects the action with the maximum Q-value as the best action for the current state. However, the action Q-values output by the network are inaccurate estimates. Continuously updating the network parameters based on these estimates can lead the model to learn biased representations of the observed features and forget prior label knowledge.

To address the knowledge forgetting issue in the original DQN, a regularized loss function has been designed. The loss function during model training consists of two parts: first, the mean squared error loss of deep reinforcement learning, which aims to make the Q-network's estimated Q-values closer to the true Q-values; second, the regularization term, which encourages the Q-network's predicted classification results to align more with prior knowledge. The definition of the regularized loss function is shown in Equation (7).
 \begin{equation}
	L_i\left(\theta_i\right)=L_i^1\left(\theta_i\right)+\lambda L_i^2\left(\theta_i\right)
	\label{equ7}
\end{equation}

Where $\lambda$ is a hyperparameter that balances the relative contributions of the mean squared error loss from deep reinforcement learning and the regularization term. This way of defining the loss function enables the agent to continuously improve its strategy under the deep reinforcement learning mechanism (MSE loss $L^1$) while preserving prior knowledge (regularization term $L^2$). It ensures that the model does not deviate from the correct trajectory even when making "wrong estimations." The specific definitions of $L_i^1\left(\theta_i\right)$ and $L_i^2\left(\theta_i\right)$ will be elaborated on below.
\subsubsection{MSE loss of DRL method}
The agent adopts a Temporal Difference (TD) based approach to learn the optimal action-value function, which doesn't require multiple complete experiments but updates the Q-values of the current state using the Q-value of the next state. To address the instability issues caused by simultaneously acquiring Q-values and updating the Q-network, a target network with delayed updates is introduced.

For computational efficiency, we didn't compute all the expectations in Equation (2). Instead, we optimized the loss function using stochastic gradient descent, replacing the expectations with single samples. The definition of the loss function is as shown in Equation (8).
\begin{equation}
	L_i^1\left(\theta_i\right)=\left(r+\gamma\max_{a^\prime}{Q\left(s^\prime,a^\prime;\theta_i^-\right)}-Q\left(s,a;\theta_i\right)\right)^2
	\label{equ8}
\end{equation}

During training, the model updates the target network parameters every $K$ steps, setting $\theta_i^-=\theta_{i-1}$. The aforementioned loss function comprises both the predicted Q-value and the target Q-value. For a given state s and action a, the Q-network outputs the predicted Q-value as $Q(s,a;\theta_i)$. The target Q-value is obtained from the Bellman equation, representing the long-term reward obtainable after taking action $a$ in the current state as
$ r+\gamma\max_{a^\prime}{Q\left(s^\prime,a^\prime;\theta_i^-\right)}$
, where r is the immediate reward provided by the environment based on state $s$ and action $a$, $s^\prime$ is the next state entered by the agent after taking action $a$, and $Q\left(s^\prime,a^\prime;\theta_i^-\right)$ is the Q-value output by the target network for the optimal action $a^\prime$ in the next state $s^\prime$.

In the end, the MSE loss is employed to gauge the disparity between the predicted Q-values and the target Q-values. By minimizing the loss function $L_i^1\left(\theta_i\right)$ and updating the parameters of the Q-network, it gradually converges towards the optimal Q-value function. This enables the agent to achieve better policy selection and higher performance.
\subsubsection{Regularization term}
The MSE loss function in the original DQN is used to measure the difference between predicted values and target values, serving as a performance metric to guide the intelligent agent in updating parameters in the correct direction. It treats the output of the Q-network as the predicted value and the output of the target Q-network as the target value. However, because the target Q-network is also derived from the Q-network, both the predicted Q-values and the target Q-values are essentially potentially inaccurate neural network estimates. These inaccuracies can lead the agent to take incorrect actions, resulting in behaviors that contradict prior knowledge and leading to cases of knowledge forgetting. In other words, the MSE loss function may not be able to completely and accurately guide the behavior of the agent.

In order to prevent the agent from making erroneous decisions due to inaccurate predicted values compared to target values, it is desired that the agent receives guidance from known prior knowledge during parameter updates, rather than solely relying on estimated information. The DML strategy proposed by Zhang et al. \cite{zhang2018deep} utilizes a dual loss function, where the student network ensemble, while learning from each other, is guided not only by the KL divergence loss function but also by the supervised cross-entropy loss function. Similarly, to enable the intelligent agent to receive supervisory signals from prior knowledge while being driven by RL, we incorporate partially labeled samples during agent optimization and adds a regularization term on top of the original MSE loss function, as defined in Equation (9).
\begin{equation}
	L_i^2\left(\theta_i\right)=-\frac{1}{M}\sum_{k=1,s_k\in\mathcal{D}_l}^My_k\cdot log\widehat{y_k}+\left(1-y_k\right)\cdot log\left(1-\widehat{y_k}\right)
	\label{equ9}
\end{equation}

Where ${M}$ is the batch size used for the regularization term, consisting of an equal number of normal and anomalous labeled samples selected from the labeled dataset $\mathcal{D}_l$. The selected samples are denoted as $s_k$, where $s_k\in\mathcal{D}_l$. $y_k$ is the true label of sample $s_k$, and $\widehat{y_k}$ is the predicted label for sample $s_k$, obtained by passing through $Q(s_k,a;\theta_i)$ and then processed through the softmax function.

This regularization term measures the discrepancy between the agent's predicted classes and the prior labels, allowing the agent to retain portions that align with the prior information when updating its parameters. Even in the face of misestimations by the double Q-network, the agent is still able to distinguish between normal and anomalous instances. To some extent, it suppresses significant changes in critical parameters of the deep reinforcement learning model, preventing the agent from taking actions that contradict prior knowledge and guiding it to learn in a direction that does not deviate from the normal trajectory.
\subsection{Algorithm Description and Analysis}
\subsubsection{The Algorithm of DQNLog}
The training process of DQNLog is depicted in Algorithm 1. First, initialize the Q-network and the target Q-network with the same weight parameters, and initialize the experience reply memory with a size of $U_{size}$. Then we train DQNLog for $n\_episodes$ episodes. For each episode, first randomly select a state $s_1$ from $\mathcal{D}_u$, and then perform training for $n\_steps$ steps. In Steps 6-8, the agent employs the $\varepsilon-greedy$ exploration strategy. It randomly selects an action from 
$\left\{a^0,a^1\right\}$ with probability $\varepsilon$; otherwise, it selects the action $t$ with the maximum Q-value. As the Q-value function improves its accuracy, 
$\varepsilon$ linearly decreases as the total time steps increase. In Steps 9-10, after the agent selects an action, the environment randomly returns state $s_{t+1}$ from 
$\mathcal{D}_l$ with probability $p$, otherwise it selects the state \(s_{t+1}\) from a random subset \(S \subset \mathcal{D}_u\) based on \(g_u\), which corresponds to the state \(s_t\) that is farthest/closest. In Steps 11-13, the environment provides a combined reward, \(r_t\), which is the sum of an external reward, \(r_t^e\), based on \(h\left(s_t,a_t\right)\), and an internal reward, \(r_t^i\), based on \(f\left(s_t\right)\). Afterwards, the obtained transition \(\left(s_t,a_t,r_t,s_{t+1}\right)\) is stored in \(\mathcal{U}\). During Steps 15-21, the Q-value function is updated. Additionally, every K steps, the Q-network resets the target Q-network.
\begin{algorithm}
	\caption{Training DQNLog}
	\begin{algorithmic}[1]
		\REQUIRE $\mathcal{D}=\left\{\mathcal{D}_l,\mathcal{D}_u\right\}$
		\ENSURE $Q\left(s,a;\theta^\ast\right)$
		\STATE Randomly initialize the parameters of both the Q-network and the target Q network, setting  $\theta^\prime=\theta$
		\STATE Initialize the size of the stored experience $ \mathcal{U}$ to $\mathcal{U}_{size}$
		\FOR{j = 1 to $n\_episodes$}
		\STATE Initial first state $s_1$ randomly from $\mathcal{D}_u$;
		\FOR{t = 1 to $n\_steps$}
		\STATE With probability $\varepsilon$ select a random action $a_t$ from $\mathcal{A}=\left\{a^0,a^1\right\}$
		\STATE Otherwise select $a_t=\arg\max_a{Q\left(s_t,a;\theta\right)}$
		\STATE $\varepsilon$ decays by $anneal\_rate$ per step;
		\STATE With probability $p$ the environment returns $s_{t+1}$ randomly from $\mathcal{D}_l$
		\STATE Otherwise returns $s_{t+1}$ from $S\subset\mathcal{D}_u$ based on $g_u$
		\STATE Calculate extrinsic reward $r_t^e$ based on $h\left(s_t,a_t\right)$
		\STATE Calculate intrinsic reward $r_t^i$ based on $f\left(s_t\right)$
		\STATE Calculate reward $r_t= r_t^e+r_t^i$;
		\STATE Store transition $\left(s_t,a_t,r_t,s_{t+1}\right)$ in $\mathcal{U}$
		\STATE Sample random minibatch of transitions $\left(s_j,a_j,r_j,s_{j+1}\right)$ from $\mathcal{U}$ 
		\STATE Set $y_j$=$
		\begin{cases}
			r_j, \quad \text{if episode terminates at this step}\\
			r_j +\gamma\max_a'{Q\left(s_{j+1},a^\prime;\theta^\prime\right)}, \quad \text{otherwise}\\
		\end{cases}$
		\STATE Calculate ${loss}_1=\left(y_j-Q\left(s_j,a_j;\theta\right)\right)^2$
		\STATE Randomly sample $\frac{M}{2}$ normal and anomaly samples $\left(s_k,y_k\right)$ separately from $\mathcal{D}_l$
		\STATE Calculate ${loss}_2=-y_k\cdot log\widehat{y_k}+\left(1-y_k\right)\cdot log(1-\widehat{y_k})$
		\STATE Calculate ${loss=loss}_1+{loss}_2$
		\STATE Perform a gradient descent step on loss with respect to the Q network weights $\theta$ 
		\STATE update $\theta^\prime=\theta$ every K steps
		\ENDFOR
		\ENDFOR
		\RETURN Q
	\end{algorithmic}
\end{algorithm}
\subsubsection{The Analysis of DQNLog}
The log anomaly detection model based on deep reinforcement learning achieves anomaly detection by approximating the optimal value-action function $Q(s,a;\theta^*)$. $Q(s,a;\theta^* )$ is the Q-network obtained after training with the parameter values $\theta^*$. For a given state $\widetilde{s}$, the Q-network outputs the estimated Q-values corresponding to taking action $a^0$ or $a^1$. Since taking action $a^1$ corresponds to the agent marking the current state $\widetilde{s}$ as anomaly, the model output $Q(\widetilde{s},a^1;\theta^*)$ can be regarded as the anomaly assessment score for state $\widetilde{s}$. To illustrate the rationale of considering $Q(\widetilde{s},a^1;\theta^*)$ as an anomaly score under the designed joint reward function, an analysis is provided:

Assuming $\pi$ is the policy derived from the Q function, then under the given policy $\pi$, for a given state $\widetilde{s}$, the expected return $q_\pi(\widetilde{s},a^1)$ obtained by the agent taking action $a^1$ is defined as shown in Equation (10).
\begin{equation}
	q_\pi(\widetilde{s},a^1)=\mathbb{E}_\pi\left[\sum_{n=t}^{\propto}{\gamma^nr_{t+n+1}}|\widetilde{s},a^1]\right]
	\label{equ10}
\end{equation}

Assuming $\widetilde{s}^{la}$ represents a known anomalous sample in the labeled set $\mathcal{D}_l$, $\widetilde{s}^{ln}$ represents a known normal sample in the labeled set $\mathcal{D}_l$, $\widetilde{s}^{ua}$  represents an anomalous sample in the unlabeled set $\mathcal{D}_u$, and $\widetilde{s}^{un}$ represents a normal sample in the unlabeled set $\mathcal{D}_u$. Then, when the agent takes action $a^1$, the external reward value given by the external reward function $h$ satisfies $h(\widetilde{s}^{la},a^1 )>h(\widetilde{s}^{ua},a^1 )=h(\widetilde{s}^{un},a^1 )>h(\widetilde{s}^{ln},,a^1 )$; the internal reward value given by the internal reward function $f$ satisfies $f(\widetilde{s}^{la},\theta^* )\approx f(\widetilde{s}^{ua},\theta^* )>f(\widetilde{s}^{un},\theta^* ) \approx f(\widetilde{s}^{ln},\theta^* )$. Since the joint reward value $r_t$ in Equation (6) is the sum of the external reward value and the internal reward value, for states $\widetilde{s}^{la},\widetilde{s}^{ln},\widetilde{s}^{ua},\widetilde{s}^{un}$, under the same policy $\pi$, the expected returns of the agent taking action $a^1$ satisfy $q_\pi(\widetilde{s}^{la},a^1 )>q_\pi(\widetilde{s}^{ua},a^1 )>q_\pi(\widetilde{s}^{un},a^1 )>q_\pi(\widetilde{s}^{ln},a^1 )$.

Therefore, when the agent is continuously trained to a convergent state and can effectively approximate the optimal action-value function, the relationship of anomaly scores output by the Q-network satisfies: $Q(\widetilde{s}^{la},a^1;\theta^* )>Q(\widetilde{s}^{ua},a^1;\theta^* )>Q(\widetilde{s}^{un},a^1;\theta^* )>Q(\widetilde{s}^{ln},a^1;\theta^* )$. This indicates that states with higher anomaly scores, $\widetilde{s}^{la},\widetilde{s}^{ua}$, which are labeled anomalous samples and unlabeled anomalous samples respectively, are the anomalies that the model needs to detect.
\section{EXPERIMENTS}
\subsection{Datasets}
We evaluated DQNLog using three widely utilized log datasets available on Loghub\cite{zhu2023loghub}, including the HDFS dataset, the BGL dataset, and the Thunderbird dataset. These datasets consist of real-world collected data, either manually annotated by system administrators or automatically generated with alert labels by the systems. A brief overview of each dataset is provided below:

(1)  \textbf{HDFS}\cite{xu2009detecting}: The HDFS dataset is sourced from the Hadoop Distributed File System(HDFS). It is generated by running MapReduce jobs on over 200 Amazon EC2 nodes and labeled by Hadoop domain experts. In total, 11,172,157 log messages were generated from 29 events, which can be grouped into 16,838 log sequences based on $block\_id$, with approximately $2.9\%$ indicating system anomalies.

(2) \textbf{BGL}\cite{oliner2007supercomputers}:The BGL dataset was generated by the Blue Gene/L(BGL) supercomputer at Lawrence Livermore National Laboratory. It was manually labeled by BGL domain experts. It consists of 4,747,963 log messages from 376 events, with 348,460 flagged as abnormal $(7.34\%)$. Log messages in the BGL dataset can be grouped by $node\_id$.

(3) \textbf{Thunderbird}\cite{xu2009detecting}: The Thunderbird dataset was generated by a Thunderbird microprocessor supercomputer deployed at Sandia National Laboratories(SNL) and manually labeled as normal or anomalous. It is a large dataset with over 200 million log messages. Among the more than one million consecutive log data collected for time calculation, 4,934 $(0.49\%)$ were identified as abnormal\cite{li2023did}. The log messages in the Thunderbird dataset can be grouped into sliding windows.

Table \ref{tab:table1} presents the comprehensive statistics for these three datasets, including the column "$\#$Log Keys" which represents the total count of distinct log templates extracted using the Drain\cite{he2017drain} technique.
\begin{table}[!t]
	\caption{Statistics of datasets
	\label{tab:table1}}
	\centering
	\begin{tabular}{cccc}
		\toprule
		\textbf{Dataset}&$\#$\textbf{Log Message}&$\#$\textbf{Anomalies}&$\#$\textbf{Log Keys}\\
		\hline
		\texttt{HDFS} & 11,172,157& 16,838(blocks)&47\\
		\texttt{BGL}& 4,747,963 & 348,460(logs) &346\\
		\texttt{Thunderbird}& 20,000,000& 4934(logs) &815\\
		\bottomrule
	\end{tabular}
\end{table}
\subsection{Implementation setup}
We conduct all the experiments on a Linux server with Intel(R) Xeon(R) Platinum 8352V 2.10GHz CPU,120GB memory,RTX 4090 with 24GB GPU memory and operating system version is Ubuntu 18.04. PyTorch was used as the deep learning framework, and the implementation was based on the Python language, utilizing the open-source Keras-rl reinforcement learning library.

DQNLog defaults to updating the parameters of the Q network after 5 episodes of pre-warming, with each episode consisting of 2000 steps. The target Q network is updated every 5 episodes. We use the Adam optimizer with a learning rate of 0.001. The size of the experience reply memory U is 100000, and the size of the mini-batch subset S is 1000. The discount factor $\gamma$ is 0.99. The hidden layer size of the Q network is 128, and the batch size ${M}$ is 32. When using the $\varepsilon-greedy$ greedy strategy, the exploration rate $\varepsilon$ is initialized to 1 and linearly decreases to 0.1 as the training progresses, where it then remains constant.
\subsection{Evaluation Metrics}
To assess the performance of DQNLog, we categorize the classification outcomes as TP, TN, FP, and FN, and employ precision, recall, and F1-score as evaluation metrics. The definitions and formulas for these three metrics are as follows:

Precision: The ratio of true positive results to the total number of log sequences predicted as anomalous. A higher precision indicates a greater proportion of actual anomalous log sequences among the predicted ones. The precision formula is: $Precision = TP / (TP + FP)$.

Recall: The ratio of true anomalous log sequences correctly identified as anomalous. A higher recall indicates a greater proportion of detected anomalous log sequences. The recall formula is: $Recall = TP / (TP + FN)$.

F1-Score: The harmonic mean of precision and recall, where higher values indicate better model performance. The F1-Score formula is: $F1-Score = (2 * Precision * Recall) / (Precision + Recall)$.

In the field of anomaly detection, it is common for normal samples to significantly outnumber anomalous samples, resulting in highly imbalanced data. Consequently, models tend to learn the characteristics of normal classes more effectively, while their ability to recognize anomalous classes is relatively weak. In this scenario, precision can effectively measure the model's accuracy in predicting anomalous samples. Meanwhile, given the abundance of unlabeled data, the model's comprehensive prediction of anomalous samples in the entire unlabeled dataset requires recall for assessment. Taking both aspects into account, the F1 score, as a combined measure of recall and precision, garners more attention.

\subsection{Compared Approaches}
To validate the effectiveness of the log anomaly detection method proposed in this paper, several mainstream models in the field of log anomaly detection were selected as baseline methods for experimental comparison. The details are as follows:

(1)LogClustering\cite{lin2016log}: Hierarchical clustering of log sequences, considering event weights, selects centroids as representative log sequences per cluster. Anomalies in log sequences are determined by comparing their distances to all centroids to classify them as normal or abnormal.

(2)PLELog\cite{yang2021semi}: The dimensionality-reduced log semantic vectors are clustered using the HDBSCAN algorithm. Similar vectors are grouped together, and unmarked log sequences are assigned probability labels based on known normal labels within each cluster. These probability labels indicate the likelihoods of log sequence categories, rather than providing definite labels. Build an attention-based GRU neural network model for anomaly detection, leveraging the labeled training set estimated via probability labels. 

(3)LogEncoder\cite{qi2023logencoder}: It combines single-class and contrastive learning objectives to distinguish between normal and abnormal log sequences. Using an attention-based model, it preserves log context, learns sequence features, and maps them onto a hypersphere. Anomalies are detected based on the distance from log sequences to the hypersphere's center.

(4)DeepLog\cite{du2017deeplog}: It parses log data into sequential vectors and parameter value vectors. Sequential vectors represent log templates by sequence IDs, while parameter value vectors correspond to the parameters of each template. Initially, sequential vectors are fed into an LSTM model to detect abnormal execution paths. Then, separate LSTM networks are built for each template to identify parameter value anomalies using parameter value vectors. Log sequences showing abnormal execution paths or parameter values are flagged as anomalies.

(5)LogAnomaly\cite{meng2019loganomaly}: It parses logs into sequential vectors and quantitative vectors, using an LSTM model to predict the next log event. If the actual next log event deviates from the predicted result, it is flagged as an anomaly.
\subsection{Results and Analysis}
For software systems, there is always a large amount of easily accessible unlabeled data and a small amount of labeled data. To validate the effectiveness of the method in this scenario, 10000 log sequences were extracted from each of three datasets. 80\% of these sequences form the training set $\mathcal{D}$, which includes a small labeled dataset $\mathcal{D}_l$ and a large unlabeled dataset $\mathcal{D}_u$, with a data ratio of $\mathcal{D}_l:\mathcal{D}_u$ being $3:7$. The remaining 20\% constitutes the test set. To ensure that the proportion of anomalies in the dataset is close to real-world scenarios, the anomaly contamination rates in the three datasets are set to 3\%, 7\%, and 1\% respectively. The experimental data selection is shown in Table \ref{tab:table2}.
\begin{table*}
	\caption{Selection of experimental data
	\label{tab:table2}}
	\centering
	\begin{tabular}{cccccc}
		\toprule
		Dataset&$\mathcal{D}_l$&$\mathcal{D}_u$&$\mathcal{D}_l:\mathcal{D}_u$&Training Data&Testing Data\\
		\hline
		\texttt{HDFS} & 2400(72 anomalies)& 5600(168 anomalies)& \multirow{3}{*} {3:7} &\multirow{3}{*}{8000}& \multirow{3}{*}{2000}\\
		\texttt{BGL}& 2400(168 anomalies) & 5600(392 anomalies) \\
		\texttt{Thunderbird}& 2400(24 anomalies)& 5600(56 anomalies)\\
		\bottomrule
	\end{tabular}
\end{table*}

To objectively evaluate the model's anomaly detection capability in scenarios with a small amount of labeled samples and a large amount of unlabeled samples, accuracy, recall, and F1-score were used as evaluation metrics. The performance was compared against several mainstream models in the field of log anomaly detection. The experimental results are shown in Table \ref{tab:table3}.
\begin{table}
	\caption{Experimental results}
	\label{tab:table3}
	\begin{tabular}{c|l|ccc}
		\toprule
		\textbf{Dataset}&\textbf{Method}&\textbf{Precision}&\textbf{Recall}&\textbf{F1-score}\\
		\midrule
		\multirow{3}{*}{\textbf{HDFS}} 
		 & LogClustering & 100\% & 78.30\% & 87.90\% \\
		& PLELog & 64.20\% & 87.20\% & 73.95\%\\
		& LogEncoder & 94.50\% & 94.30\% & 94.40\%\\
		& DeepLog  & 90.87\% & 89.42\% & 90.14\%\\
		& LogAnomaly & 92.50\%	&90.70\% &91.59\% \\
		\cline{2-5}
		& DQNLog & 98.14\% & 94.05\% & 96.06\%\\
		\hline
		\multirow{3}{*}{\textbf{BGL}} 
		& LogClustering & 72.20\% & 21.70\% & 33.30\% \\
		& PLELog & 53.36\% & 98.30\% & 69.17\%\\
		& LogEncoder & 93.50\% & 87.60\% & 89.60\%\\
		& DeepLog & 84.74\% & 80.78\% & 82.73\%\\
		& LogAnomaly & 90.87\% & 88.02\% & 89.42\%\\
		\cline{2-5}
		& DQNLog & 98.27\% & 86.99\% & 92.29\%\\
		\hline
		\multirow{3}{*}{\textbf{Thunderbird}} 
		& LogClustering & 88.90\% & 76.7\% & 82.35\% \\
		& PLELog & 68.60\% & 87.50\% & 76.46\% \\
		& LogEncoder & 87.50\% & 90.20\% & 88.90\%\\
		& DeepLog & 91.45\% & 91.02\% & 91.23\%\\
		& LogAnomaly &92.13\%&91.24\%&91.68\% \\
		\cline{2-5}
		& DQNLog & 94.64\% & 94.64\% & 94.64\%\\
		\bottomrule
	\end{tabular}
\end{table}

The results indicate that compared to the benchmarked semi-supervised and unsupervised methods, DQNLog achieves higher accuracy while maintaining good recall, resulting in the best overall F1 score performance. This is because DQNLog leverages labeled prior knowledge to learn features of known anomalies. Furthermore, driven by deep reinforcement learning reward signals, it explores anomalies in unlabeled datasets effectively. In contrast, PLELog and LogEncoder only train on normal instances, completely disregarding known anomalies, leading to a high recall but low precision by frequently misclassifying normal log sequences as anomalous, ultimately resulting in lower F1 scores. Methods like DeepLog, LogAnomalyr treat each sample equally during training, often favoring the majority class in imbalanced data scenarios, thus achieving higher accuracy but lower recall and consequently lower F1 scores. Traditional machine learning approaches like LogClustering generally perform poorly in log anomaly detection due to their reliance solely on log count vectors, which ignore inter-log correlations and semantic relationships, making them inadequate for handling dynamic changes in log data.

Meanwhile, observations across different datasets such as HDFS, BGL, and Thunderbird show that various anomaly detection models consistently perform better on the HDFS dataset. This is because the BGL and Thunderbird datasets contain a much larger number of templates compared to the HDFS dataset, with many log templates being rare or even absent in the training set. As a result, models struggle to effectively learn the characteristics of rare templates, leading to an overall performance decrease.Specifically, it can be noted that on the BGL dataset, all anomaly detection models achieve lower recall rates. This is primarily due to the long time spans of session windows in the BGL dataset, where multiple types of anomalies may occur within the same session. Models find it challenging to effectively learn multiple types of anomalies, resulting in decreased recall rates.
\subsection{Ablation Study}
\begin{figure}[!t]
	\centering
	\includegraphics[width=3.5in]{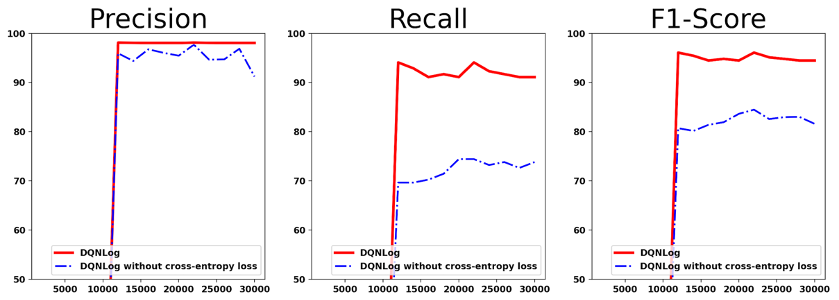}
	\caption{The comparison results between DQNLog and $DQNLog_{no\_cross}$ on the HDFS dataset}
	\label{cross}
\end{figure}

To validate the effectiveness of the main components in DQNLog, comparisons were made between DQNLog and its variants. The two types of defined ablation variants are as follows:

(1)$DQNLog_{no\_cross}$: a variant in which the loss function does not include the regularization term. Figure \ref{cross} shows the comparison results between DQNLog and $DQNLog_{no\_cross}$ on the HDFS dataset. The results indicate that adding the regularization term constrains the model with prior information when exploring anomalies, contributing to the overall performance improvement of DQNLog.

More specifically, compared to $DQNLog_{no\_cross}$, DQNLog reaches convergence quickly, whereas $DQNLog_{no\_cross}$ continues to fluctuate and performs significantly worse across all three evaluation metrics. This instability in $DQNLog_{no\_cross}$ arises because it relies solely on experience samples from the experience reply memory during training, which may be inaccurate as they are generated by the Q network. In contrast, DQNLog incorporates a regularization term that leverages prior information from log sequence labels during gradient descent updates, in addition to Q network estimates. This helps DQNLog maintain accurate parameter updates, even when the estimates are imprecise.

\begin{figure}[!t]
	\centering
	\includegraphics[width=3.5in]{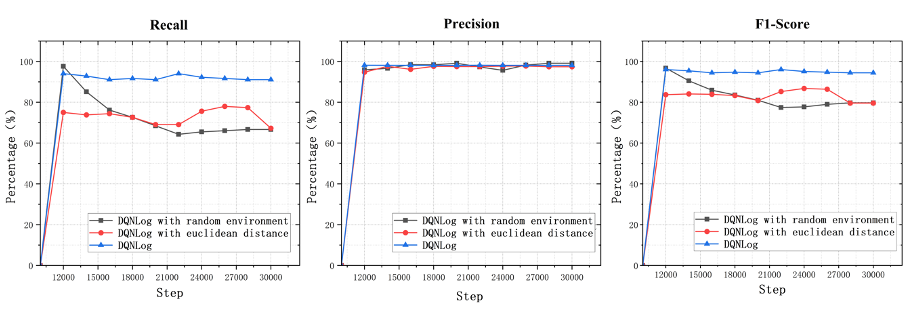}
	\caption{The comparative results of DQNLog, $DQNLog_{random\_env}$ and $DQNLog_{euc\_env}$ on the HDFS dataset}
	\label{env}
\end{figure}

(2)$DQNLog_{random\_env}$ and $DQNLog_{euc\_env}$: $DQNLog_{random\_env}$ operates in a random environment with stochastic state transition functions, while $DQNLog_{euc\_env}$ operates in a biased anomalous environment but utilizes Euclidean distance to calculate the similarity of log vectors. Figure \ref{env} illustrates the comparative results of DQNLog, $DQNLog_{random\_env}$, and $DQNLog_{euc\_env}$ on the HDFS dataset. The results indicate that the biased anomalous state transition function based on cosine similarity increases the agent's probability of effectively exploring potential anomalies in large-scale unlabeled datasets. This approach alleviates the issue of class imbalance present in the dataset.

More specifically, DQNLog performs more stably compared to $DQNLog_{random\_env}$ and $DQNLog_{euc\_env}$. When the number of unlabeled data is large and the actual proportion of anomalies is small, the model can overcome the influence of the majority "normal" class and tends to learn log sequences with more "anomalous" characteristics. This is because cosine similarity considers both the length and direction of vectors, which better reflects the similarity of logs compared to random functions and Euclidean distance. As a result, the next state returned by the environment is more likely to be "anomalous", allowing the model to learn more anomalous information, significantly improving recall and F1 scores.
\subsection{Hyperparameter analysis}
\begin{figure}[!t]
	\centering
	\includegraphics[width=3.5in]{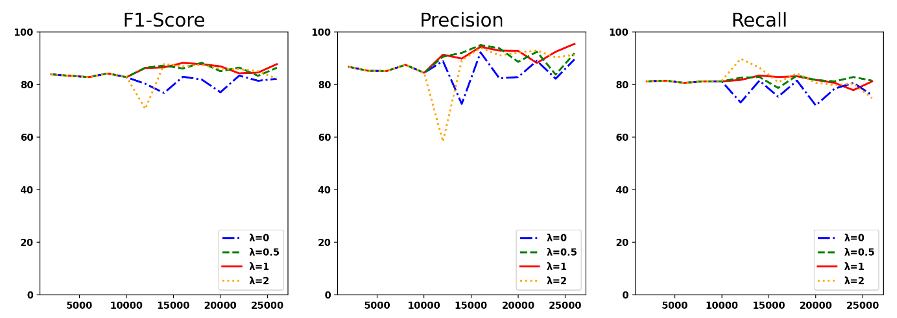}
	\caption{Experimental results for different values of $\lambda$}
	\label{reward}
\end{figure}

To explore the impact of related hyperparameter values on the performance of DQNLog, we varied the values of the hyperparameters to determine the optimal parameters for the model. The investigated hyperparameters include:

(1)The value of the regularization term coefficient $\lambda$ in the loss function. Figure \ref{reward} illustrates the variation of precision, recall, and F1 score during the training process on the HDFS dataset under different values of $\lambda$.

The results indicate that when $\lambda = 2$, the model shows significant fluctuations in precision, recall, and F1 score during training. In contrast, when $\lambda = 0.5$ or $\lambda = 1$, the metrics of the model exhibit relatively stable variations. This may be attributed to an excessive amplification of the regularization term, which overly preserves prior knowledge and results in unstable performance changes. Specifically, increasing $\lambda$ from 0.5 to 1 improves the overall performance slightly, allowing the model to leverage the information brought by prior knowledge to a greater extent.

Therefore, it is considered that when the value of the regularization term coefficient $\lambda$ is in the range $[0.5, 1]$, the model can achieve better detection performance. Excessively large hyperparameter values can cause model fluctuations, while excessively small values cause the model to ignore the prior knowledge of log sequences, ultimately leading to the capture of less information by the model.

(2)The penalty values for false positives (FP) and false negatives (FN) in the reward function, $r_3$ and $r_4$. Table \ref{tab:table4} demonstrates the detection performance of DQNLog on the HDFS dataset as the false positive (FP) penalty varies, with a fixed false negative (FN) penalty. Table \ref{tab:table5} illustrates the detection performance of DQNLog on the HDFS dataset as the false negative (FN) penalty varies, with a fixed false positive (FP) penalty.
\begin{table}[!t]
	\caption{Fix FN, vary FP}
	\label{tab:table4}
	\centering
	\begin{tabular}{cccc}
		\toprule
		\ &\textbf{Precision}&\textbf{Recall}&\textbf{F1-score}\\
		\hline
		FN -1, FP -0.2 & 98.07\%& 92.26\%&95.08\%\\
		FN -1, FP -0.3& 98.10\% & 91.67\% &94.78\%\\
		FN -1, FP -0.4& 98.11\%& 91.07\% &94.44\%\\
		FN -1, FP -0.5& 98.11\%& 91.07\% &94.44\%\\
		FN -1, FP -0.6& 98.13\%& 90.48\% &94.12\%\\
		\bottomrule
	\end{tabular}
\end{table}

The results indicate that as the penalty for false positives increases, the model's recall gradually decreases, while precision gradually increases. This suggests that a model with a higher false positive penalty tends to be more "cautious" in predicting anomalies. The model will only classify a current log sequence as anomalous when it is highly confident that the sequence differs significantly from the majority of log sequences, thereby improving precision. However, this may also lead to an increase in missed actual anomalous logs, resulting in a decrease in the model's recall.
\begin{table}[!t]
	\caption{Fix FP, vary FN}
	\label{tab:table5}
	\centering
	\begin{tabular}{cccc}
		\toprule
		\ &\textbf{Precision}&\textbf{Recall}&\textbf{F1-score}\\
	\hline
	FN -1, FP -0.4 & 98.11\%& 91.07\%&94.44\%\\
	FN -1.2, FP -0.4& 98.10\% & 91.67\% &94.78\%\\
	FN -1.5, FP -0.4& 98.09\%& 92.26\% &95.10\%\\
	FN -2, FP -0.4& 97.53\%& 92.26\% &94.77\%\\
	\bottomrule
\end{tabular}
\end{table}

The results indicate that as the penalty for false negatives increases, the model's precision gradually decreases, while recall gradually increases. This suggests that a model with a higher false negative penalty tends to be more "bold" in predicting anomalies. It encourages the system to learn from more potentially anomalous samples to avoid missing a significant number of actual anomalies, thereby achieving higher recall. However, it also increases the likelihood of normal log sequences being falsely classified as anomalies, leading to a decrease in precision.
\section{CONCLUSION}
To address challenges such as insufficient labeled data, underutilization of unlabeled data, and imbalance between normal and anomaly class data in existing log anomaly detection methods, this paper proposes a deep reinforcement learning-based approach named DQNLog for software log anomaly detection. The method aims to leverage a small amount of labeled data and large-scale unlabeled data effectively. Specifically, it designs a state transition function biased towards anomalies based on cosine similarity, incorporates a joint reward function using external and internal rewards, introduces a regularization term in the loss function, and utilizes a Bi-LSTM network with attention mechanisms as the agent architecture. Evaluation on three widely used log datasets confirms the effectiveness of DQNLog.

Future work includes, but is not limited to, exploring methods with higher computational efficiency and greater accuracy in identifying anomalies as the intrinsic reward mechanism to optimize model performance. Additionally, enhancing knowledge during log embedding using domain-specific data to refine the representation of feature vectors more accurately.

\newpage

\vfill

\end{document}